\documentstyle[editedvolume]{crckapb} 
\input{psfig.sty}

\begin{opening}

\title{ISO FAR-IR SPECTROSCOPY OF IR-BRIGHT GALAXIES AND ULIRGS}

\author{J. FISCHER}
\author{M.L. LUHMAN}
\institute{Naval Research Laboratory, Washington, DC, USA}

\author{S. SATYAPAL}
\author{M.A. GREENHOUSE}
\institute{NASA Goddard Space Flight Center, Greenbelt, MD, USA}

\author{G.J. STACEY}
\author{C.M. BRADFORD}
\institute{Cornell University, Astronomy Department, Ithaca, NY, USA}

\author{S.D. LORD}
\author{J.R. BRAUHER}
\institute{California Institute of Technology, IPAC, Pasadena, CA, USA}

\author{S.J. UNGER}
\author{P.E. CLEGG}
\institute{Queen Mary \& Westfield College, Univ. of London, London, UK}

\author{H.A. SMITH}
\author{G. MELNICK}
\institute{Harvard-Smithsonian, CfA, Cambridge, MA, USA}

\author{J.W. COLBERT}
\author{M.A. MALKAN}
\institute{Univ. of California, Dept. of Astronomy, Los Angeles, CA, USA}

\author{L. SPINOGLIO}
\institute{Istituto di Fisica dello Spazio Interplanetario-CNR, Roma, Italy}

\author{P. COX}
\institute{Institut d'Astronomie Spatiale, Orsay, France}

\author{V. HARVEY}
\author{J.-P SUTER}
\author{V. STRELNITSKI}
\institute{Maria Mitchell Observatory, Nantucket, MA, USA}
\end{opening}

\begin{document}

\begin{abstract}

Based on far-infrared spectroscopy of a small sample of nearby
infrared-bright and ultraluminous infrared galaxies (ULIRGs) with the
ISO Long Wavelength Spectrometer \footnote{Based on observations with
ISO, an ESA project with instruments funded by ESA Member States
(especially the PI contries:  France, Germany, the Netherlands and the
United Kingdom) with the participation of ISAS and NASA.} we find a
dramatic progression in ionic/atomic fine-structure emission line and
molecular/atomic absorption line characteristics in these galaxies
extending from strong [O~III]52,88\,$\mu$m and [N~III]57\,$\mu$m line
emission to detection of only faint [C~II]158\,$\mu$m line emission
from gas in photodissociation regions in the ULIRGs.  The molecular
absorption spectra show varying excitation as well, extending from
galaxies in which the molecular population mainly occupies the ground
state to galaxies in which there is significant population in higher
levels.  In the case of the prototypical ULIRG, the merger galaxy Arp
220, the spectrum is dominated by absorption lines of OH, H$_2$O, CH,
and [O~I].  Low [O~III]88\,$\mu$m line flux relative to the integrated
far-infrared flux correlates with low excitation and does not appear to
be due to far-infrared extinction or to density effects.  A progression
toward soft radiation fields or very dusty H~II regions may explain
these effects.

Key~words: infrared-bright galaxies --- ultraluminous galaxies --- 
far-infrared spectra --- starbursts --- interstellar medium.

\end{abstract}

\section{Introduction}

In order to compare the evolutionary status, energetics, obscuration,
and physical conditions of the nuclear regions of ULIRGs with those of
less luminous infrared-bright galaxies with minimal sensitivity to
extinction, we have used the grating mode of the ISO Long Wavelength
Spectrometer (LWS) \cite{clegg96} to carry out (1) a full far-infrared
spectral survey of a small sample of nearby IR-bright galaxies
including the ULIRGs Arp~220 and Mkn~231 and (2) a fine-structure line
survey of more distant galaxies, including a survey of ULIRGs in the
[C~II]158\,$\mu$m fine-structure line.  The observations allow us to
analyze the dust continuum, to put constraints on the ionization
parameters and the intensity of the radiation as it impinges upon the
surrounding neutral clouds of gas and dust, and ultimately on the
nature of the source(s) of luminosity.  Detailed analyses of the
spectra of the individual galaxies are presented elsewhere
(\citeauthor{fis96}, \citeyear{fis96}, \citeyear{fis97};
\citeauthor{colb99}, \citeyear{colb99}; \citeauthor{saty99},
\citeyear{saty99}; \citeauthor{lord99}, \citeyear{lord99};
\citeauthor{ung99}, \citeyear{ung99}; \citeauthor{brad99},
\citeyear{brad99}; \citeauthor{har99}, \citeyear{har99}; and
\citeauthor{spin99}, \citeyear{spin99}).   Here we present a
comparative overview of the observational results.

\section{The LWS full spectra of infrared-bright galaxies}

\begin{figure}[!hp]
    \psfig{file=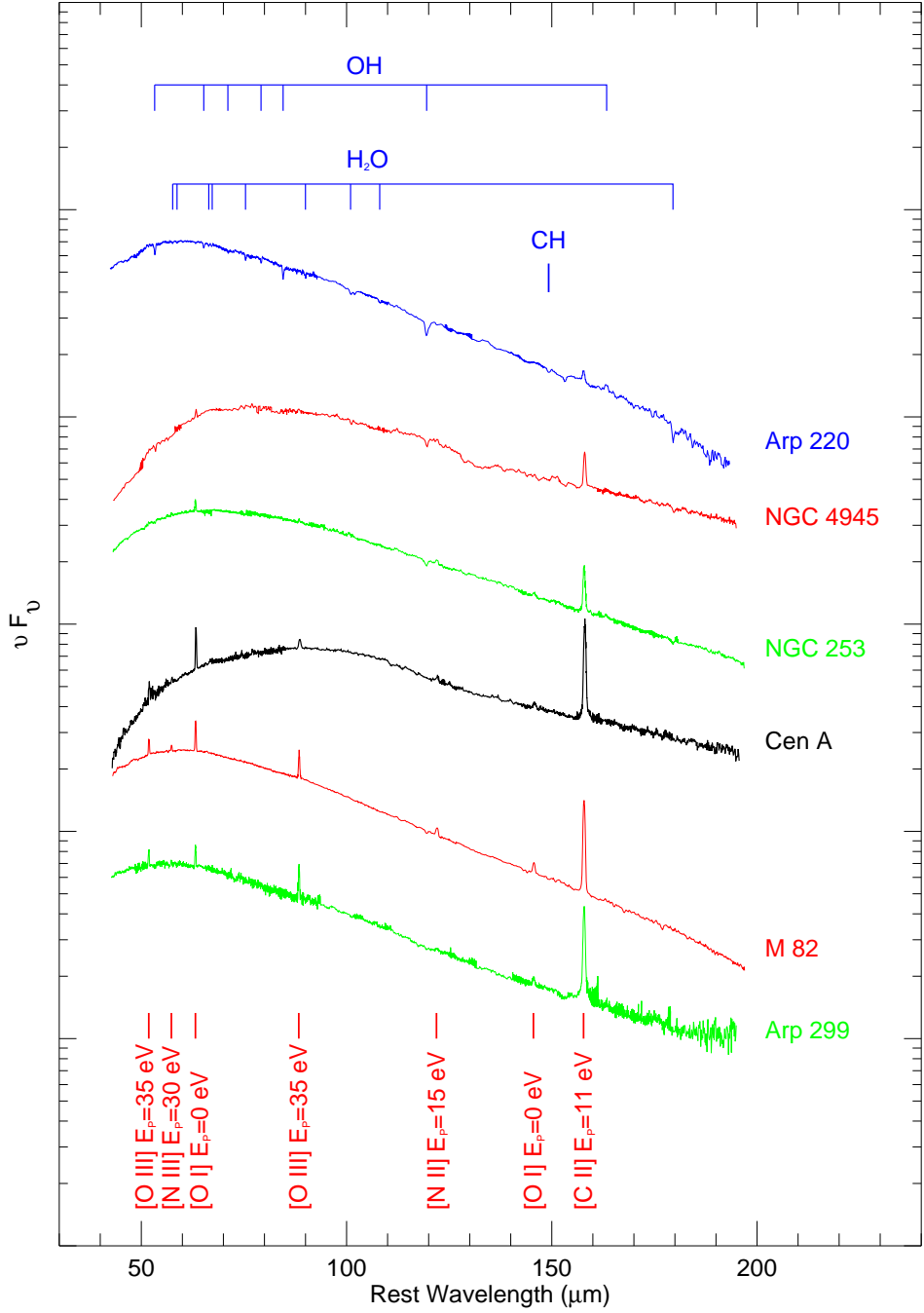,width=\textwidth}
\caption{\em The full ISO Long Wavelength Spectrometer spectra of six 
IR-bright galaxies.  The spectra have been shifted and ordered vertically
according to apparent excitation (Fischer et~al. 1999) and are not in
order of relative luminosity or brightness.} 
\label{fig:firseq}
\end{figure}

We present full, high signal-to-noise LWS spectra of six
infrared-bright galaxies in Figure~\ref{fig:firseq}.  The LWS aperture
is $\sim$ 75" \cite{swin98} and the spectral resolution is $\lambda
/\Delta \lambda \sim $ 200 in grating mode.  The spectra are presented
in sequence extending from strong [O~III]52,88\,$\mu$m and
[N~III]57\,$\mu$m fine-structure line emission in the galaxies Arp~299
and M~82, to only faint [C~II]158 $\mu$m line emission from gas in
photo-dissociation regions in the prototypical ultraluminous galaxy
Arp~220.  The far-infrared spectrum of the ULIRG Arp~220 is dominated
by absorption lines of OH, H$_{2}$O, CH, and [O~I].  Intermediate in
the sequence are Cen~A, NGC~253, and NGC~4945, showing weak [O~III] and
[N~III] lines while their PDR emission lines remain moderately strong.
Interestingly, the strength and richness of the molecular absorption
spectra is anti-correlated with the equivalent widths of the
fine-structure emission lines.  For example, M~82 shows faint OH
absorption from the ground level at 119\,$\mu$m \cite{colb99}, while
NGC~253 shows absorption from the ground-state in three cross-ladder
transitions and an emission line cascade at 79\,$\mu$m and 163\,$\mu$m
\cite{brad99}.  In NGC~4945 and Arp~220, OH absorption from both ground
and excited rotational levels is present (\citeauthor{lord99},
\citeyear{lord99}; \citeauthor{fis97}, \citeyear{fis97}).  In Arp~220,
although the existence of a downward cascade is suggested by the
presence of emission at 163\,$\mu$m, absorption from rotational levels
as high as 416 K and 305 K above the ground state is seen for OH and
H$_2$O, respectively, and the [O~I]63\,$\mu$m line is seen in
absorption.  Although the location of the excited molecules is not
certain, OH and H$_2$O are expected to exist in abundance in dense
photo-dissociation regions (PDRs) \cite{stern95}, where they could be
excited radiatively by the far-infrared emission from warm dust.

It is of interest to compare the far-infrared spectra of the
archetypical Seyfert 2 galaxy NGC~1068 \cite{spin99} and that of the
Galactic Center \cite{white99} with the spectra presented in
Figure~\ref{fig:firseq}.  The equivalent widths of the far-infrared
fine-structure line emission in NGC~1068 resemble those in the
starburst galaxy M~82.  In addition to its Seyfert 2 nucleus, NGC~1068
hosts a starburst in its circumnuclear ring, that is possibly as young
as the youngest clusters in M~82 \cite{dav98}.  This starburst may be
responsible for much of the far-infrared emission, as suggested by
\citeauthor{tel84} \shortcite{tel84}.  A notable difference is that in
NGC~1068 the OH lines are observed in emission suggesting unique
excitation conditions possibly related to its Seyfert 2 nucleus (see
discussion in \citeauthor{spin99}, \citeyear{spin99}), while in M~82 OH
is observed in absorption \cite{colb99}.  We note here that for Cen~A
(\citeauthor{ung99}, \citeyear{ung99}; Figure~\ref{fig:firseq}), also
known to harbor an AGN, the weakness of the far-infrared [O~III] lines
may indicate that the AGN is not the dominant source powering the
far-infrared luminosity.  The far-infrared spectrum of the Galactic
Center \cite{white99} would fall toward the upper end of the sequence
shown in Figure~\ref{fig:firseq}, with an added emission line component
due to warm, perhaps shock-excited, neutral gas.

\section{Parameterization of the far-infrared spectral sequence 
of IR-bright galaxies}

The sequence shown in Figure~\ref{fig:firseq} may be caused by
variation of many parameters, but it is of interest to examine whether
a single parameter or evolutionary effect can play the dominant role in
the progression, and in particular to try to understand what conditions
cause the ultraluminous galaxies to appear at the extreme end of the 
sequence.

\begin{figure}
  \psfig{file=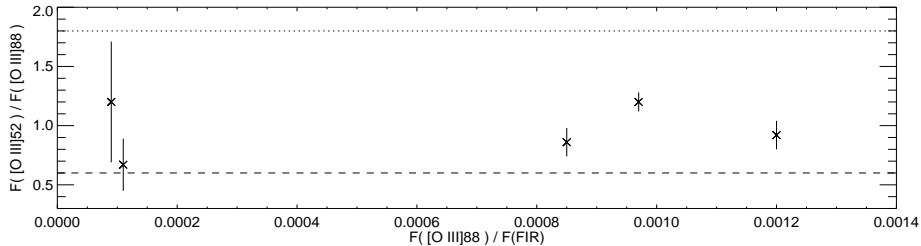,width=\textwidth}
\caption{\em The [O~III]52\,$\mu$m/[O~III]88\,$\mu$m line ratio versus the 
[O~III]88\,$\mu$m line to integrated far-infrared continuum flux ratio for
the sample galaxies.  The dashed and dotted lines show the [O~III] line
ratio in the low density limit ($\leq$ 100 $cm^{-3}$) and for an
electron density of 500 cm$^{-3}$, respectively (Fischer et al. 1999).} 
\label{fig:oxyrat}
\end{figure}

In Figure ~\ref{fig:oxyrat} we plot the temperature-insensitive
[O~III]52/[O~III]88 line ratio as a function of the [O~III]88/$F_{FIR}$
ratio for the galaxies in which [O~III] line emission was detected in
Figure~\ref{fig:firseq}.  To within the uncertainties no clear
dependence was found for our small sample and all of the measured
[O~III] line ratios fall within the range 0.6 - 1.2, consistent with
electron densities between 100 - 500 cm$^{-3}$.  These results suggest
that neither density nor far-infrared differential extinction between
52 and 88\,$\mu$m appears to be the \emph{single dominant parameter} in
the observed sequence \cite{fis99}.  This is consistent with previous
extinction estimates.  Despite the inferred high column density of dust
corresponding to A$_v$ $\geq$ 1000, the estimated extinction \emph{to
the ionized gas} in Arp~220 is A$_v$ $\sim$ 25-50 (\citeauthor{fis97},
\citeyear{fis97}; \citeauthor{gen98}, \citeyear{gen98}).

\begin{figure}
  \psfig{file=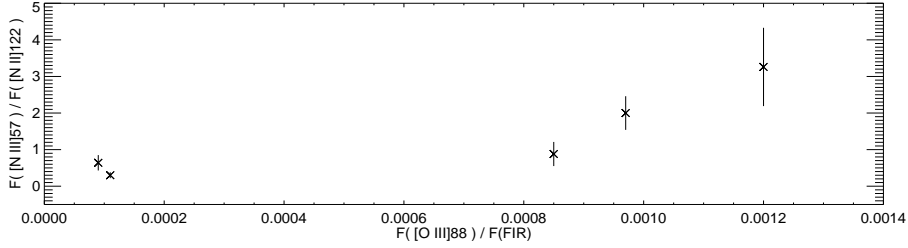,width=\textwidth}
\caption{\em As in Figure 2 for the [N~III]57\,$\mu$m/[N~II]122\,$\mu$m 
line ratio.} 
\label{fig:nit32}
\end{figure}
\begin{figure}
  \psfig{file=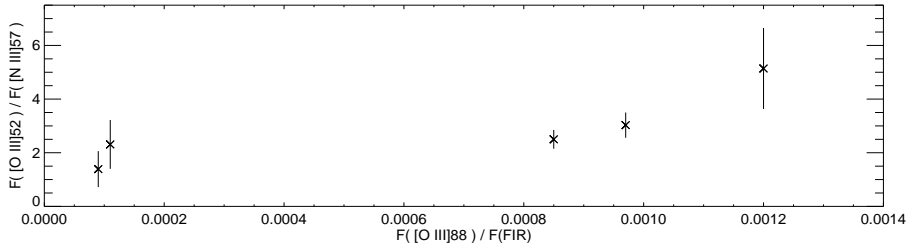,width=\textwidth}
\caption{\em As in Figure 2 for the [O~III]52\,$\mu$m/[N~III]57\,$\mu$m
line ratio.}
\label{fig:oxy3nit3}
\end{figure}

In Figures~\ref{fig:nit32} and ~\ref{fig:oxy3nit3} we plot the
[N~III]57/[N~II]122 and [O~III]52/[N~III]57 line ratios as a function
of the [O~III]88/$F_{FIR}$ ratio for galaxies from
Figure~\ref{fig:firseq} (where these lines were detected).  The
excitation potentials of [N~II], [N~III], and [O~III] are 14.5, 29.6,
and 35.1 eV, respectively.  Thus for constant metallicity, both
Figures~\ref{fig:nit32} and \ref{fig:oxy3nit3} indicate that with
progression to low relative emission line strength is a progression to
lower excitation.  Figure~\ref{fig:oxy3nit3} strengthens our conclusion
that far-infrared extinction is not responsible for the apparent
excitation effects.  The ionization parameter, defined as the ratio of
ionizing photons to hydrogen atoms at the inner face of the cloud,
plays a key role in determining ionization structure of clouds
surrounding a source of ionizing radiation.  It is equal to
Q$^\prime$(H)/4$\pi r^2n_Hc$, where Q$^\prime$(H) is the Lyman
continuum rate absorbed by the gas and $n_H$ is the hydrogen density at
the inner radius, $r$, of the cloud.  Thus if density effects alone do
not explain the sequence, effects such as larger inner cloud radii due
to stellar winds or lower Q$^\prime$(H)/$L_{Bol}$ due to dust within
the HII regions or softer radiation fields may be responsible for the
apparent excitation progression.  If the latter is the case, and if
starbursts are the source of the excitation, then an aging starburst or
one with an IMF with a low upper mass limit could be present in the
ultraluminous galaxies and other low excitation galaxies.  Soft
radiation fields or dusty H~II regions may explain the presence of
ubiquitous molecular material in close proximity to the nuclear regions
of these galaxies and the prominent molecular absorption lines.  It is
difficult however, to reconcile the aging starburst interpretation with
the high luminosity of the ultraluminous galaxies, since older
starbursts have lower luminosities than their younger counterparts.

\section{The far-infrared spectra of ULIRGs}

The far-infrared spectrum of the second brightest ultraluminous galaxy
Mkn~231 \cite{har99} is surprisingly similar to that of Arp~220 (to
within the achieved signal-to-noise ratio).  It is dominated by
OH absorption, with similar OH absorption line ratios, and only
faint PDR line emission is present.  A single component absorption
layer is inconsistent with the observed line ratios in these galaxies
and fluorescent components do not alleviate the problem.  The observed
OH line ratios probably result from independent absorption and emission
components \cite{sut98}.  Based on the mid-infrared spectra of a sample
of nearby ULIRGs, \citeauthor{gen98} \shortcite{gen98} infer that
Mkn~231 has a strong AGN component while the far-infrared luminosity of
Arp~220 is powered by a starburst.  Thus the similarity of the
far-infrared spectra of these two ultraluminous galaxies is somewhat
surprising.

The ultraluminous galaxies have lower [C~II]158\,$\mu$m line to
far-infrared flux ratios than in normal and less luminous IR-bright
galaxies by an order of magnitude (\citeauthor{luh98},
\citeyear{luh98}; \citeyear{luh99}).  This has been interpreted as an
indication of a lower value of the average interstellar radiation
field $\langle$$G_o$$\rangle$ in Arp~220, where the upper limit for the
[O~I]145/[C~II]158 emission line ratio is unexpectedly low
(\citeauthor{fis97}, \citeyear{fis97}; \citeauthor{luh98},
\citeyear{luh98}).  Implicit in this interpretation is the assumption
that the [O~I]145\,$\mu$m upper limit is not affected by
self-absorption, a reasonable assumption since the lower level of the
[O~I]145\,$\mu$m line is 228 K above the ground state.  On the other
hand, if self-absorption is responsible for the apparent faintness of
the [O~I]145\,$\mu$m line, then very high values of $\langle$$G_o$$\rangle$
are possible, as has been suggested by \citeauthor{mal98}
\shortcite{mal98} for a small percentage of their sample of normal
galaxies.  A plausible explanation for both low ionization parameters
and low values of $\langle$$G_o$$\rangle$ is dusty H~II regions.  Low
ionization parameters can be consistent with high $\langle$$G_o$$\rangle$
if molecular clouds surround very compact H~II regions.

\section*{Acknowledgments}

This work was supported in part by the Office of Naval Research and the
NASA ISO grant program.  We appreciate the skill and dedication of the LWS
instrument and data analysis teams at Vilspa, RAL, and IPAC.

\end{document}